\documentclass[aps,prd,twocolumn,superscriptaddress,showpacs]{revtex4}
\usepackage{graphicx}
\usepackage{amsmath}
\usepackage{amsfonts}
\usepackage{amssymb}

\begin{document}
\title{Upper bounds on the photon mass}
\author{Antonio Accioly}\email{accioly@cbpf.br}
\affiliation{Laborat\'{o}rio de F\'{\i}sica Experimental (LAFEX), Centro Brasileiro de Pesquisas F\'{i}sicas (CBPF), Rua Dr. Xavier Sigaud 150, Urca, 22290-180, Rio de Janeiro, RJ, Brazil}
\affiliation{Group of Field Theory from First Principles,  S\~{a}o Paulo State University (UNESP), Rua Dr. Bento Teobaldo Ferraz 271, Bl. II-Barra Funda, 01140-070 S\~{a}o Paulo, SP, Brazil}
\affiliation{Instituto de F\'{\i}sica Te\'{o}rica (IFT), S\~{a}o Paulo State University (UNESP), Rua Dr. Bento Teobaldo Ferraz 271, Bl. II-Barra Funda, 01140-070 S\~{a}o Paulo, SP, Brazil}

\author{Jos\'{e} Helay\"{e}l-Neto}\email{helayel@cbpf.br}
\affiliation{Laborat\'{o}rio de F\'{\i}sica Experimental (LAFEX), Centro Brasileiro de Pesquisas F\'{i}sicas (CBPF), Rua Dr. Xavier Sigaud 150, Urca, 22290-180, Rio de Janeiro, RJ, Brazil}
\affiliation{Group of Field Theory from First Principles,  S\~{a}o Paulo State University (UNESP), Rua Dr. Bento Teobaldo Ferraz 271, Bl. II-Barra Funda, 01140-070 S\~{a}o Paulo, SP, Brazil}

\author{Eslley Scatena}\email{scatena@ift.unesp.br}
\affiliation{Instituto de F\'{\i}sica Te\'{o}rica (IFT), S\~{a}o Paulo State University (UNESP), Rua Dr. Bento Teobaldo Ferraz 271, Bl. II-Barra Funda, 01140-070 S\~{a}o Paulo, SP, Brazil}
\affiliation{Group of Field Theory from First Principles,  S\~{a}o Paulo State University (UNESP), Rua Dr. Bento Teobaldo Ferraz 271, Bl. II-Barra Funda, 01140-070 S\~{a}o Paulo, SP, Brazil}

\begin{abstract}
The effects of a nonzero photon rest mass can be incorporated into electromagnetism in a simple way using the Proca equations. In this vein, two interesting implications regarding the possible existence  of a massive photon in nature, i.e., tiny alterations in the known values of both the  anomalous magnetic moment of the electron and   the  gravitational deflection of electromagnetic radiation, are utilized to set upper limits on its mass. The bounds obtained are not as stringent as those recently found; nonetheless, they are comparable to other existing bounds and bring new elements to the issue of  restricting the  photon mass.
 \end{abstract}
\maketitle
\section{Introduction}
In general, systems of heavy vector bosons are non-renormalizable.  There are, however,  two important exceptions to this rule: (i) gauge theories with spontaneous symmetry breakdown, and (ii) Abelian theories with neutral vectorial bosons coupled to conserved currents \cite {1}. The latter, i.e., the `conserved current models', contain at least one massive boson, whose source is conserved. These systems  can be constructed through the following general prescription \cite{2}: 

\begin{itemize}
\item Begin with a Lagrangian which is invariant under a nonsemisimple group of local gauge transformations (i.e., a group containing an invariant Abelian subgroup).
\item Arrange for spontaneous symmetry breaking (if any) such that the vacuum expectation value of the scalar field is invariant under  at least one invariant (single-parameter) Abelian subgroup (thus, at  this stage the  corresponding Abelian vector is massless and coupled to a conserved current).
\item Add (in the $R$ gauge) an arbitrary mass term for the same Abelian vector.
\end{itemize}

\noindent Massive electrodynamics (or, Proca electrodynamics), i.e, the electrodynamics that can be embedded into the standard $SU(2)\times U(1)$ model and in which the photon has a small mass, is the simplest system of this type, besides being the most straightforward extension of standard QED. Indeed, Proca's electromagnetic field theory can be constructed in a unique way by adding a mass term  to the Lagrangian for the electromagnetic field, namely,
 
\begin{eqnarray}
{\cal L} = -\frac{1}{4}F_{\mu \nu}^2 -J_\mu A^\mu + \frac{1}{2} m^2 A_\mu^2,
\end{eqnarray}

\noindent where $F_{\mu \nu } \;(= \partial_\mu A_\nu - \partial_\nu A_\mu)$ is the field strength, and $J_\mu$ is the (electric) current. The parameter $m$ can be interpreted as the photon rest mass. In this spirit, the characteristic scaling length $m^{-1}$ becomes the reduced Compton wavelength of the photon, which is the effective range of the electromagnetic  interaction.  

 Massive QED is not only simpler theoretically than the standard theory \cite{3}, it also provides a fairly solid framework for analyzing (through the Proca equations) the  far-reaching implications the existence of a massive photon would have for physics. Actually, some of these possible effects, such as variation of the speed of light, deviation in the behavior of static electromagnetic fields and longitudinal electromagnetic radiation, have been thoroughly  studied by means of a number of different approaches over the past several decades \cite{4,5,6}. It is worth mentioning that both the Aharonov-Bohm  and the  Aharonov-Casher effects are present in massive QED. The former was analyzed by  Boulware and Deser \cite {7} who showed that it reduces  smoothly to the original result, while the latter was studied by Fuchs \cite{8}. Nonetheless, the system of `Maxwell + photon mass + magnetic charge' equations is not consistent \cite{3,4}. 

Interestingly enough, the possibility of a nonzero photon mass remains, as it was pointed out by Adelberger, Dvali, and Gruzinov \cite{9}, one of the most important issues in physics, as it would shed a new light on some fundamental questions, such as charge conservation, charge quantization, the possibility of charged black holes and magnetic monopoles. We also remark that the popular view that gauge invariance implies a zero photon mass is not correct. In reality, a minimal dynamics obeying gauge invariance, i.e., the Maxwell action, does imply zero photon mass; nevertheless, by enlarging the dynamics, for instance, by adding another field interacting with the photon field, both gauge invariance and nonzero mass can be accommodated simultaneously \cite{6}.

The purpose of this paper is to set upper bounds on the photon mass supposing that it  is described by Proca electrodynamics. To accomplish this goal we shall analyze two interesting but not yet explored consequences of the  possible existence of a massive photon in nature: the very small alteration in the usual anomalous magnetic moment of the electron and  the  tiny change in the ordinary gravitational deflection of the electromagnetic radiation. These issues are analyzed  in detail in Secs. II and III, respectively. To conclude, a discussion about the order of  magnitude of the bounds estimated  in the  aforementioned sections, is presented in Sec. IV.

In our conventions $\hbar = c =1$, and  the signature is (+ - - -). 

\section{A QUANTUM BOUND}
As is well-known, QED predicts the anomalous magnetic moment of the electron correctly to ten decimal places. Therefore, it is perfectly reasonable that we use this astonishing result, one of the great triumphs of QED, to estimate a quantum bound on the photon mass. How can we do that? By computing the anomalous magnetic moment of the electron  to order $\alpha$, where $\alpha$ is the fine structure constant, in the framework of massive QED and expanding afterward the result in powers of $(\frac{m}{\mu})^2$, where $m$ and $\mu$ are, respectively, the photon and the electron masses. The first term of this expansion must necessarily coincide with that calculated by Schwinger in 1948 \cite{10}, while the second one is the most important correction related to the parameter $m$ of massive QED. Now, taking into account that the latter must be less than $10^{-10}$ (the theoretical result predicted by QED for the anomalous magnetic moment of the electron \cite {11} agrees in $1$ part in $10^{10}$ with the experimental one \cite {12}),  we promptly find an upper bound for the photon mass.

Let us then perform the computations. We begin by recalling that the anomalous magnetic moment  of the electron stems from the vertex correction for the scattering of the electron by an external field, as it is shown in Fig. 1.

\begin{figure}[!hbp]
\begin{center}
\includegraphics[scale=0.58]{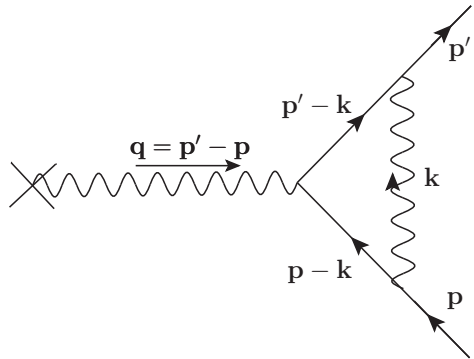}
\end{center}
\caption{\small Vertex correction for electron  scattering by  an external  field.\label{fig1}}
\end{figure}

For an electron scattered by an external static magnetic field and in limit ${\bf q}\rightarrow {\bf 0}$, the gyromagnetic ratio is given by \cite{13}
$$g =2[ 1 + F_2(0)].$$

\noindent The form factor of the electron, $F_2 (0)$, corresponds to a shift in the $g-$factor, usually quoted in the form $F_2(0)\equiv \frac{g-2}{2}$, and yields the anomalous magnetic moment of the electron. 

On the other hand, from the quadratic part of Lagrangian (1) we immediately obtain the propagator for the massive QED, namely,

\begin{eqnarray}
D_{\mu \nu} = - \frac{\eta_{\mu \nu}}{k^2 - m^2} + \frac{k_\mu k_\nu}{m^2(k^2 - m^2)}.
\end{eqnarray}

\noindent By employing this expression in the calculation of the diagram in Fig. 1, it can be shown that

\begin{eqnarray}
F_2 (0) = \frac{\alpha}{\pi}\int_0^\infty d\alpha_1 d\alpha_2 d \alpha_3 \delta (1 - \Sigma \alpha_i)\frac{\alpha_1(\alpha _2 + \alpha_3)}{(\alpha _2 + \alpha_3)^2 + \lambda^2 \alpha_1}, \nonumber
\end{eqnarray} 
 
\noindent where $\lambda^2 \equiv (\frac{m}{\mu})^2$. We remark that the term $\frac{k_\mu k_\nu}{m^2(k^2- m^2)}$ that appears in Eq. (2) was omitted {\it ab initio} from the calculations concerning $F_2(0)$ because the  propagator for the massive photon  always occurs coupled to conserved currents.

In order to avoid unnecessary algebraic computations as far as the evaluation of $F_2(0)$   is concerned, we rewrite this expression as follows:

$$F_2(0)= X-Y,$$

\noindent where

\begin{eqnarray}
X\equiv \frac{\alpha}{\pi}\int_0^\infty d\alpha_1 d\alpha_2 d \alpha_3 \delta (1 - \Sigma \alpha_i)\frac{\alpha_1}{\alpha _2 + \alpha_3},
\end{eqnarray}

\begin{eqnarray}
Y \equiv  &=& \frac{\alpha}{\pi}\int_0^\infty d\alpha_1 d\alpha_2 d \alpha_3 \delta (1 - \Sigma \alpha_i)\left[\frac{\alpha_1}{\alpha _2 + \alpha_3} \nonumber  \right. \\&&- \left. \frac{\alpha_1 (\alpha_2 + \alpha_3)}{(\alpha_2 + \alpha_3)^2 + \lambda^2 \alpha_1}  \right].
\end{eqnarray}

Integrating the expression (3) first with respect to $\alpha_3$ and subsequently with respect to $\alpha_2$ gives

\begin{eqnarray}
X &=& \frac{\alpha}{\pi}\int_0^1 d\alpha_1 \int_0^{1-\alpha_1} d\alpha_2 \frac{ \alpha_1}{1- \alpha_1 } \nonumber   \\&=&  \frac{\alpha  }{2\pi}.
\end{eqnarray}  
 
  Similarly,  the expression (4) yields

\begin{eqnarray}
Y  &=& \frac{\alpha}{\pi}\int_0^1d\alpha_1 \frac{\lambda^2 \alpha_1^2}{(1 - \alpha_1)^2 + \lambda^2 \alpha_1 } \nonumber   \\&=& \frac{\alpha}{\pi} \left[\lambda^2 - ( \lambda^4 - 2\lambda^2) \ln \lambda  \right. \nonumber \\&&+ \left.  \frac{\lambda^5 - 4\lambda^3 + 2\lambda}{\sqrt{4 - \lambda^2}} \arctan \frac{\sqrt{4- \lambda^2}}{\lambda} \right].
\end{eqnarray}

From (5) and (6) we get

\begin{eqnarray}
F_2(0)  &=& \frac{\alpha}{\pi}\left[ \frac{1}{2}- \lambda^2( 1 + 2 \ln \lambda) + \lambda^4 \ln \lambda \right. \nonumber \\&&+ \left.   \frac{4\lambda^3 - 2\lambda - \lambda^5}{\sqrt{4 - \lambda^2}} \arctan \frac{\sqrt{4- \lambda^2}}{\lambda} \right].
\end{eqnarray}

Recalling that $\lambda \ll 1 $, we arrive at the conclusion that

\begin{eqnarray}
F_2(0)  &\approx& \frac{\alpha}{2\pi}\left[ 1- \pi \left(\frac{m}{\mu}\right) -\left( \frac{3}{2} + 4 \ln \left(\frac{m}{\mu}\right)\right)\left( \frac{m}{\mu}\right)^2 \right. \nonumber \\&&+ \left.   {\cal \large O} \left(\left( \frac{m}{\mu}\right)^3\right)\right].
\end{eqnarray}

\noindent As we have already commented, the first term in the above equation is equal to that calculated by Schwinger in 1948 (since then $F_2(0)$ has been calculated to order $\alpha^8$ for QED), while the second one is the most important correction concerning the parameter $m$ of  massive QED. Since theory and experiment agree within errors to $\sim$ $1$ in $10^{10}$ for $F_2(0)$, we promptly obtain

\begin{eqnarray}
\frac{\pi m}{\mu} < 10^{-10},
\end{eqnarray}

\noindent implying $m < 1.6 \times 10^{-10} MeV.$

Recently, the measurement of the anomalous magnetic moment of the muon  reached the fabulous relative precision of 0.5 ppm  \cite{14,15}. Accordingly, it would   be interesting to find another quantum bound on the photon mass using this phenomenon and  make afterward a comparison with the bound estimated  via the electron.  Now, taking into account that for the muon \cite{16}

$$F_2^{(\mathrm{exp})}(0) - F_2^{(\mathrm{SM})}(0) = (295 \pm 88) \times 10^{-11},$$

\noindent where $F_2^{(\mathrm{SM})}(0)$ denotes the prediction of the standard model \cite{17}, we find $m < 3.4 \times 10^{-7} MeV$, three orders of magnitude higher than the bound derived from the anomalous magnetic moment of the electron. Consequently, we shall not consider this bound in our discussions.

\section{A GRAVITATIONAL BOUND}
It is a generally acknowledged fact that the gravitational deflection of light by  the sun can be measured more accurately at radio wavelengths with interferometry techniques than  at visible wavelengths with available optical techniques \cite{18}. Indeed, at present the  Very Long Baseline Interferometry (VLBI) is the most accurate technique we have at our disposal  for measuring radio-wave gravitational deflection \cite{19,20,21}. The gravitational bending, in turn, is  one of the most impressive predictions of general relativity. In addition, the recent measurements of the gravitational bending of radio waves using the VLBI have improved considerable on the previous  results in the gravitational bending experiments near the solar limb \cite{22}.    Accordingly, we shall use these results to estimate an upper limit on the photon mass. To do that we need, in first place, the unpolarized differential cross section for the scattering of a massive photon (described by Proca's electrodynamics) by an external weak gravitational field. On the other hand, it was  recently shown that the unpolarized differential cross sections for the gravitational scattering of  different quantum particles are spin dependent \cite{23} (See Table I). Nonetheless,  for small angles, the cross sections for the massive (massless) particles are one and the same, regardless of the spin. In fact, when the spin is `switched off', i.e., for small angles ($\theta \ll 1$), it  is fairly straightforward to see from Table I that for $m=0$,  $\frac{d\sigma}{d \Omega} \approx \frac{16 G^2 M^2}{\theta^4}$, while for $m\neq 0$, $\frac{d\sigma}{d \Omega} \approx \frac{16 G^2 M^2}{\theta^4} \left( 1 + \frac{\lambda}{2} \right)^2.$  In short, for small angles the results of Table I are in perfect agreement with those  predicted by Einstein's geometrical theory. Consequently, the  differential cross section we are searching for is independent of the spin of the massive particle and can be written as  
\begin{eqnarray}
\frac{d\sigma}{d \Omega} = \frac{16 G^2 M^2}{\theta^4} \left( 1 + \frac{\lambda}{2} \right)^2.
\end{eqnarray}

The above differential cross section can be related to a classical trajectory with  impact parameter $b$ via the relation 
\begin{eqnarray}
bdb = - \frac{d\sigma}{d\Omega} \theta d\theta.
\end{eqnarray} 

From (10) and (11), we arrive at the conclusion that

\begin{eqnarray}
\theta = \frac{4MG}{b}\left(\frac{1 - \frac{m^2}{2E^2}}{1 - \frac{m^2}{E^2}}\right),
\end{eqnarray}

\noindent which in the ultrarelativistic limit, i.e., $E \gg m$,  reduces to

\begin{eqnarray}
\theta &=& \theta_{\mathrm E}\left(1 + \frac{m^2}{2E^2}\right), \nonumber \\ &=&  \theta_{\mathrm E}\left( 1 + \frac{m^2}{8 \pi^2 \nu^2}\right),
\end{eqnarray}

\noindent where $E$ and $\nu$ are, respectively, the energy and the frequency of the ingoing massive photon, and $ \theta_{\mathrm E} \equiv \frac{4MG}{b}$.

\begin{table}[here]
\caption{\label{table1}Unpolarized differential cross sections for the scattering of different quantum particles by an external weak gravitational field generated by a static point particle of mass $M$. Here $m$ is the particle mass, $s$ the spin, $\theta$ the scattering angle, $G$ the Newtonian constant,  and $\lambda\equiv\frac{m^{2}}{\mathbf{p}^{2}}=\frac{1-\mathbf{v}^{2}}{\mathbf{v}^{2}}$, with $\mathbf{v}$ and $\mathbf{p}$ being the velocity and three-momentum, in this order, of the incident particle.}
\begin{ruledtabular}
\begin{tabular}{ccc}
$m$&$s$&$\frac{d\sigma}{d\Omega}$\\
\hline
\\
0 & 0 & $\Big(\frac{GM}{\sin^{2}{\frac{\theta}{2}}}\Big)^{2}$\\
\\
$\neq 0$ & 0 & $\Big(\frac{GM}{\sin^{2}{\frac{\theta}{2}}}\Big)^{2}\Big(1+\frac{\lambda}{2}\Big)^{2}$\\
\\
0 & $\frac{1}{2}$ & $\Big(\frac{GM}{\sin^{2}{\frac{\theta}{2}}}\Big)^{2}\cos^{2}{\frac{\theta}{2}}$\\
\\
$\neq 0$ & $\frac{1}{2}$ & $\Big(\frac{GM}{\sin^{2}{\frac{\theta}{2}}}\Big)^{2}\Big[\cos^{2}{\frac{\theta}{2}}+\frac{\lambda}{4}\Big(1+\lambda+3\cos^{2}{\frac{\theta}{2}}\Big)\Big]$\\
\\
0 & 1 & $\Big(\frac{GM}{\sin^{2}{\frac{\theta}{2}}}\Big)^{2}\cos^{4}{\frac{\theta}{2}}$\\
\\
$\neq 0$ & 1 & $\Big(\frac{GM}{\sin^{2}{\frac{\theta}{2}}}\Big)^{2}\Big[\frac{1}{3}+\frac{2}{3}\cos^{4}{\frac{\theta}{2}}-\frac{\lambda}{3}\Big(1-\frac{3\lambda}{4}-4\cos^{2}{\frac{\theta}{2}}\Big)\Big]$\\
\\
0 & 2 & $\Big(\frac{GM}{\sin^{2}{\frac{\theta}{2}}}\Big)^{2}\Big(\sin^{8}{\frac{\theta}{2}}+\cos^{8}{\frac{\theta}{2}}\Big)$\\ 
\end{tabular}
\end{ruledtabular}
\end{table}

The first term in the expression (13) coincides with that obtained by Einstein in 1916 by solving the equation of light propagation in the field of a static body \cite{24}, whereas the second one is the most important correction due to the mass $m$ of the massive photon. On the other hand, the angle of gravitational bending measured by the experimental groups is expressed in general trough the relation \cite{25}

\begin{eqnarray}
\theta_{\mathrm {exp}} = \frac{1 + \gamma}{2} \theta_{\mathrm E},
\end{eqnarray}

\noindent where $\gamma$ is the deflection parameter characterizing the contribution of space curvature to gravitational deflection.  From Eqs. (13) and (14), we then get

\begin{eqnarray}
\theta_{\mathrm E}  \frac{m^2}{8 \pi^2 \nu^2} < \theta_{\mathrm E} \left(1- \frac{1 + \gamma}{2}\right),
\end{eqnarray}

\noindent implying

\begin{eqnarray}
m <  2 \pi \nu \sqrt{\left|1 - \gamma\right|}.
\end{eqnarray}

 Not long ago, Fomalont {\it et al.} \cite{22} determined  the deflection parameter $\gamma = 0.9998\pm 0.0003$ (68\text{}\% confidence level), using the VLBI at $43, 23$ and $15 GHz$ to measure the solar gravitational deflection of radio waves. Their results come mainly from $43 GHz$ observations where the refraction effects of the solar corona were negligible beyond 3 degrees from the sun \cite{26}.

Using the result for the deflection parameter found by Fomalont {\it et al.} and assuming that the massive photon passing near the solar limb has a frequency $\nu = 43 GHz$ (which is perfectly justifiable in view of the argument previously provided), we conclude that $m < 3.5 \times 10^{-11} MeV$.  

We remark that Eq. (12) can also be deduced {\it \`a la Einstein}, namely, by finding an approximate solution to the geodesic equation of motion of a massive test particle in the Schwarzschild field. By adopting this approach, an expression for the angle of particle deflection by the sun was obtained to order $\left(\frac{GM}{b}\right)^3$ in Ref. 27. This kind of deduction, however, is a time-consuming work. On the other hand, Golowich, Gribosky, and Pal \cite{28}, instead of  taking the usual geometrical approach, considered the phenomenon of light bending as a quantum scattering problem. This treatment, which is not only instructive but also straightforward when the gravitational field is weak, allowed them to  easily obtain an expression for the gravitational deflection of massive particles to order $\frac{GM}{b}$. An identical result was found by Mohany, Nieves, and Pal \cite{29} using a method pioneered by Ohanian and Ruffini \cite{30}.   

At this point, some comments are in order.

\begin{itemize}
\item According to general relativity, photons are not only deflected but also delayed by the curvature of space-time produced by any mass. And more, the bending and delay are proportional to $\gamma + 1$. Consequently, time delay techniques can also be employed to set up bounds on the photon mass. It is interesting to note that a few years ago,  Bertotti, Iess, and Tortora \cite{31} reported a measurement of the frequency shift of radio photons to and from the Cassini spacecraft as they passed near the sun that led to a result for $\gamma$ which agrees with the predictions of standard general relativity with a sensitivity that approaches the level at which, theoretically, deviations are expected in some cosmological models \cite{32,33}.

\item 

 Equation (13) was derived on the assumption that the field responsible for the photon deflection is a static gravitational field. Nonetheless, as well-known, neither the sun nor the planets are at rest in the solar system. Actually, they are moving with respect to both the barycenter of the solar system and the observer. This motion will certainly bring about velocity-dependent corrections to the general-relativistic equation of the gravitational deflection of light. As a consequence, the aforementioned motion-induced correction to the gravitational deflection of light shall correlate with the correction to the photon's mass exhibited in equation (13). This fact leads us to pose an important question: Currently, is  modern technology sensitive enough to detect these tiny relativistic effects caused by the dependence of the gravitational field on  time? Kopeikin \cite{34} claims that `future gravitational light-ray deflection experiments \cite{35}, radio ranging BepiColombo experiment \cite{36}, laser ranging experiments ASTROD \cite{37} and LATOR \cite{38} will definitely reach the   
precision in measuring  $\bar{\gamma}_\mathrm{PPN},\; \bar{\beta}_\mathrm{PPN}$ and $\bar{\delta}_\mathrm{PPN}$ that is comparable with  the post-Newtonian corrections to the static time delay and to the deflection angle caused by the motion of the massive bodies in the solar system \cite{39}.' Here, deviation from general relativity are denoted with the comparative PPN parameter   $\bar{\gamma}_\mathrm{PPN}\equiv \gamma_\mathrm{PPN} - 1$, $\bar{\beta}_\mathrm{PPN} \equiv \beta_\mathrm{PPN} -1$ and $\bar{\delta}_\mathrm{PPN} \equiv \delta_\mathrm{PPN} - 1$. On the other hand, one can show, using the equation for the post-post-Newtonian time delay, $\Delta t$, which was obtained by Kopeikin by coupling  the PPN parameters with the velocity-dependent terms, that for gravitational experiments with light propagating in the field of the sun,

\begin{equation}
\Delta t \approx \left( 1 + \bar{\Gamma }\right) \ln \left( \frac{r_1 + r_2 + r_{12}}{r_1 + r_2 -r_{12}}\right),
\end{equation}

\noindent with

\begin{equation}
\bar{\Gamma} \approx \bar{\gamma}_\mathrm{PPN} - 2 \beta_\odot,
\end{equation}

\noindent where $\beta_\odot$ ($= 5.3 \times 10^{-8}$) is the solar velocity (in natural units) with respect to the barycenter of the solar system, $r_{12}$ is the coordinate distance between the emission and observation points, $r_1, \; r_2$ are radial distances to the emission and observation points, respectively. Now, noticing that the LATOR and ASTROD space missions are going to measure the $\bar{\gamma}_\mathrm{PPN}$ parameter with a precision approaching to $10^{-9}$ \cite{37,38}, we arrive at the conclusion that in the near-future, the explicit velocity-dependent correction to the static  time delay in the solar gravitational field must  apparently be taken into account. Let us then answer the question raised above. For the sake of simplicity we restrict our discussion to measurements of light bending  by the sun obtained trough VLBI techniques. Currently  the experimental groups have determined the parameter $\bar{\gamma}_\mathrm{PPN}$ using the   VLBI with  an accuracy of $ 10^{-4}$ \cite{22}. Therefore, the alluded velocity-dependent correction is too small and can be neglected in the determination of $\bar{\gamma}_\mathrm{PPN}$. Actually, the detection of so small an effect is beyond current technology.  

\item 
 Nowadays,  as we have already  pointed out, the VLBI is the most accurate technique we have at our disposal for measuring radio-wave gravitational deflection on a regular basis \cite{19,20,21}. It was only superseded by the multiple frequency Doppler-tracking of Cassini spacecraft \cite{31}. 

\item   Measuring light deflection with optical techniques may turn out more advantageous for determining the parameter $\gamma$ in a foreseeable future \cite{40}.

\end{itemize}

\section{Discussion}
We discuss now whether or not the bounds we have found could be improved. To begin with, we consider the quantum limit. A quick glance at Eq. (9) clearly shows that a better agreement between theory and experiment concerning the anomalous magnetic moment of the electron    necessarily  leads to an improvement on the quantum bound. Consequently, there is a great probability of obtaining a better   quantum  bound on the photon mass   in the foreseeable future. We analyze in the sequel  how a better limit on the photon mass might be obtained using Eq. (16). First, if the deflections measured using the VLBI could be made with greater  accuracy the value of $\sqrt{\left|1 - \gamma\right|}$ would be reduced giving, as a result, a better  gravitational estimate. According to Fomalont $et\; al.$ \cite{22}, a series of designed experiments with the VLBI could increase the accuracy of the future experiments by at least a factor of 4. Second,  if deflection measurements can be obtained at lower frequencies, while maintaining  the value of the deflection parameter $\gamma$, the gravitational bound will be improved in direct proportion to  the frequency. This point, however, needs to be dealt with carefully. In fact, as we have already mentioned, up till now  the best results obtained for the gravitational deflection via the VLBI are those that  come mainly from $43 GHz$ where the refraction effects of the solar corona are negligible beyond 3 degrees from the sun.  Incidentally, the lowest frequency  employed by the radio  astronomers was $2 GHz$. However, the measurements made at this frequency are less reliable because of the refraction effects of the solar corona. Actually, the radio astronomers use in their experiments a mixing of different frequencies but the most significant  contributions come in general from $\sim 43 GHz$. This possibility of increasing the gravitational limit is thence very limited.

Certainly, the bounds we have found  on the photon mass are  higher than the recently recommended limit published by the Particle Data Group  \cite{12}. They are nevertheless comparable to other existing bounds (See Table II) and bring new elements to the issue of restricting the photon mass. Accordingly, they do have some merits. We discuss their main qualities in the following.

\begin{table}[here]
\begin{center}
\caption{ Some upper bounds  on the photon mass obtained by measuring the dispersion in the speed of light in different ranges of the electromagnetic spectrum (in chronological order).}
\begin{tabular*}{8cm}{@{\extracolsep{\fill}}ccc}
\hline
\hline\\
\small{Author}& \small{Type of}&\small{Limits on $m$}\\
\small{(year)}& \small{measurement}& \small{($MeV$)}
\\ \\
\hline
\small{Froome}&\small{Radio-wave}&\small{$2.4\times 10^{-13}$}\\
\small{(1958)\cite{41}}&\small{interferometer}&
\\
\small{Warner {\it et al.}}&\small{Observations on Crab}&\small{$2.9 \times 10^{-14}$}\\
\small{(1969)\cite{42}}&\small{Nebula pulsar}&
\\
\small{Bay {\it et al.}}&\small{Pulsar emission}&\small{$1.7\times 10^{-19}$}\\
\small{(1972)\cite{43}}&&
\\
\small{Brown {\it et al.}}&\small{Short pulses}&\small{$7.9 \times 10^{-7}$}\\
\small{(1973)\cite{44}}&\small{radiation}&
\\
\small{Schaefer}&\small{Gamma ray bursts}& \small{$2.4\times 10^{-17}$}\\
\small{(1999)\cite{45}}&\small{(GRB980703)}&\\
&\small{Gamma ray bursts}&\small{$3.4\times 10^{-12}$}\\
&\small{(GRB930229)}&\\
\hline
\end{tabular*}
\end{center}
\end{table}

\begin{itemize}                
\item The theory adopted to describe the photon mass  has the correct limit.

\item The bounds are based on exact calculations performed in the framework of QED and general relativity, respectively; besides, the most accurate experimental data currently available have been taken as input. 

\item The conceptual approaches adopted to  estimate the bounds are new. 

\item The methods used for placing the bounds  are interesting in their own, although they do not lead to the most stringent limits. Indeed, the quantum bound is estimated using one of the most renowned predictions of QED --- the anomalous magnetic moment of the electron, while the  gravitational bound is obtained using  the properties of gravity. Essentially, the point is that a massive photon is bent in a gravitational field by a different amount than a massless photon. Thus, observations of light bending by the sun allow one to place limits on the photon mass. 

\item The bounds are essentially a measurement of the agreement between theory and experiment. Since the two limits are of the same order, they may be used to give an idea of how much the theoretical prediction deviates from the experimental result. For the quantum  and semiclassical bounds we have estimated this lower limit is $m^{-1} \sim 2cm$. Thus, the more  the value of $m^{-1}$ increases, the more the concordance between theory and experiment increases. In other words, a null mass for the photon would imply  a perfect agreement between theory   and experiment

\item Recently, Adelberger, Dvali, and Gruzinov \cite{9} questioned the validity of some bounds on the photon mass available in the literature. They claim that if $m$ arises from a Higgs effect, these limits are invalid because the Proca vector potential of the galactic magnetic field may be neutralized by vortices giving a large-scale magnetic field that is effectively Maxwellian. However, these criticisms do not apply to our computations because they  are based on the plausible assumption of large galactic vector potential; furthermore, in our case $m$ does not arise from a Higgs effect. 
\end{itemize}
Last but not least, we would like to draw the reader's attention to the article by Barton and Dombey \cite{46} in which  it is demonstrated that the Casimir effect is not sensitive to a small photon mass. To accomplish this, they showed that the contribution to the Casimir force due to the photon mass is proportional to $m^4$, being, as a consequence, negligible compared with the leading finite-mass correction to the contribution from the transverse modes. On the other hand, if the galactic magnetic field is in the Proca regime,  the very existence of the observed large-scale magnetic field gives $m \sim 10^{-26}eV$ \cite{9}. Therefore, the electron anomalous magnetic moment and the deflection of light by the sun, like the Casimir effect, are insensitive to a photon mass less than the allowed already established limits.

\begin{acknowledgments}
 The authors are very grateful to FAPERJ, CNPq, and  CAPES (Brazilian agencies) for financial support.
\end{acknowledgments}

\end{document}